\documentclass[12pt]{article}

\usepackage[a4paper,left=30mm,right=20mm,top=20mm,bottom=20mm]{geometry}
\usepackage{graphics}
\usepackage{amsmath}
\usepackage{epsfig}
\usepackage{cite}
\usepackage{xcolor}
\usepackage[unicode]{hyperref}
\title{Nuclear transverse momentum dependent gluon density at low $x$ and inclusive soft hadron production in proton-lead collisions at LHC}

\author{A.V.~Lipatov$^{1}$, G.I.~Lykasov$^{2}$, M.A.~Malyshev$^{1,3}$}

\begin{document}

\maketitle

\begin{center}
{\it $^{1}$Skobeltsyn Institute of Nuclear Physics, Lomonosov Moscow State University, 119991 Moscow, Russia}\\
{\it $^{2}$Joint Institute for Nuclear Research, 141980 Dubna, Moscow region, Russia}\\
{\it $^{3}$Moscow Aviation Institute, 125993 Moscow, Russia}\\

\end{center}

\vspace{0.5cm}

We report the results of calculations of
inclusive soft hadron production in proton-lead collisions at the LHC
in the framework of modified quark-gluon string model (QGSM)
extended to $pA$ interactions.
Our consideration involves the nuclear modification
of previously proposed transverse momentum dependent (TMD, or unintegrated) gluon density
 in a proton, which
provides a self-consistent simultaneous
description of numerous HERA and LHC data on $pp$, $ep$ and $\gamma p$ processes.
Such nuclear modification is based on well established property
of geometrical scaling from nucleons to nuclei.
Focusing on the region of small $x$ and low scales,
we obtain predictions for transverse momentum
spectra of pions and kaons at $p_T \leq 1$~GeV. Our results
are compared with recent data reported by the CMS, ATLAS and ALICE Collaborations at $\sqrt s = 5.02$~TeV.
We find that the developed approach provides a better description of low-$p_T$ data
than the predictions made by other groups.

\indent

\vspace{1.0cm}


\vspace{1.0cm}

\newpage

It is known that proton-nucleus ($pA$) collisions at high energies provide an opportunity to investigate
the role of nuclear environment in modifying binary nucleon-nucleon hard scattering cross sections.
Such modification,
observed in a number of experiments,
demolishes, of course, the naive idea of the nucleus $A$ as a system of quasi-free nucleons (see, for example, reviews\cite{NuclReview-1, NuclReview-2, NuclReview-3}).
Several physical effects are expected to induce
deviations from a simple proportionality between the measured production cross sections and
number of binary nucleon-nucleon collisions.
In fact, nuclear shadowing was reported,
indicating that the nucleus has a rather complicated nucleon structure
suppressing hadron production rates from low to moderate momenta.
There are also nuclear anti-shadowing, EMC and Fermi motion effects.
At large $x$,
nuclear effects 
are related to the smearing of the cross sections with the nuclear momentum distribution\cite{NuclearMomentumDistribution}
and nuclear binding corrections\cite{NuclearBindingCorrections}, which have kinematical origin.
At intermediate and small $x$, these effects are related to the
dynamical modification of internal parton structure,
meson exchange currents and nuclear shadowing (see, for example,\cite{NuclearShadowingApproach-1, NuclearShadowingApproach-2} and references therein).
Such effects could be investigated\footnote{Usually,
    the nuclear modification factor $R_{pA}$, defined as ratio of per-nucleon
    cross section $1/A \, \sigma_{pA}$ and proton-proton cross
    section $\sigma_{pp}$ is introduced and its behaviour in different kinematical regions is
    studied.} experimentally by measuring
the inclusive charged hadron production as function of their transverse momenta $p_T$.

Experimental data on inclusive charged hadron production in $pA$ collisions (to be precise, proton-lead ones)
have been obtained by the CMS\cite{ChargedHadron5-CMS}, ATLAS\cite{ChargedHadron5-ATLAS} and ALICE\cite{ChargedHadron5-ALICE} Collaborations at $\sqrt s = 5.02$~TeV.
Besides the studying of nuclear interaction dynamics,
these measurements provide an opportunity to constrain different models of hadron production
and contribute to understanding of basic non-perturbative dynamics in hadron collisions.
In particular, they provide a useful input for the determination of nuclear parton (gluon and quark) distribution functions (nPDFs),
which are necessary tools for theoretical description and proper interpretation (within the standard QCD factorization concept)
of $eA$, $pA$ and $AA$ processes
studied at modern (LHC, RHIC) and future colliders (FCC-he, EiC, EicC, CEPC, NICA).
The reported measurements\cite{ChargedHadron5-CMS, ChargedHadron5-ATLAS, ChargedHadron5-ALICE} are
found to be in good agreement with
predictions obtained using \textsc{epos lhc} event generator\cite{EPOS},
which is based on a quantum mechanical multiple scattering approach
with taking into account nuclear effects related to transverse momentum broadening,
parton saturation and screening. Other Monte-Carlo event generators, \textsc{ampt}\cite{AMPT}
and \textsc{hijing}\cite{HIJING}, reveal substantial deviations from the LHC
data (see \cite{ChargedHadron5-CMS} for more information).

We note, however, that the experimental data\cite{ChargedHadron5-CMS, ChargedHadron5-ATLAS, ChargedHadron5-ALICE} 
at not very large momentum transfer
can be also analyzed within the modified quark-gluon string model (QGSM)\cite{mQGSM-2,GLLZ}
extended to $pA$ collisions (see below).
In the standard QGSM\cite{QGSM-1} based on the Regge behaviour of the cross sections and
initially developed for $pp$ events, the interaction dynamics is
governed by two colorless strings formed between the quark/diquark ($q/qq$)
and diquark/quark ($qq/q$) of the colliding protons.
At their breaking, the quark-antiquark and
diquark-antidiquark pairs are created in the chromostatic QCD field
and then fragment into a final hadron $h$.
This model describes data in hadron production in $pp$ collisons
satisfactorily at large $x$. However, the standard QGSM is unable to describe
data at low $x$.
In the modified QGSM approach\cite{mQGSM-2,GLLZ}, existence of
nonperturbative (or soft) gluons in the proton,
which split into $q\bar q$ pairs and give an additional contribution to the
hadron spectrum, was suggested.
Such improvement accompanied with some
physically motivated expression\cite{LLM-2022, LLM-2024} for
the transverse momentum dependent (TMD, or unintegrated) gluon density in a proton\footnote{Transverse momentum dependent
gluon density is an essential part of the High Energy Factorization\cite{HighEnergyFactorization},
or $k_T$-factorization approach\cite{kt-factorization} (see reviews\cite{TMD-review, TMD-review-our} for more information).} leads to a good description of LHC data on soft hadron
production in $pp$ collisions in the mid-rapidity ($y \simeq 0$)
region (see\cite{LLM-small-x-fit} and references therein).
It is important that the proposed TMD gluon distribution 
provides a self-consistent
simultaneous description of numerous HERA and LHC data on $pp$, $ep$ and $\gamma p$ processes
and has been extended recently\cite{nLLM-2024} to the nuclei using the well established property of geometrical scaling\cite{GeometryScaling-1} (see also\cite{GeometryScaling-2}).
The latter is a characteristic feature of low-$x$ data on the deep inelastic lepton-proton scattering at HERA.
Main goal of our present note is to
\textcolor{violet}{extend} the developed approach\cite{mQGSM-2,GLLZ} to soft hadron production in $pA$ collisions
and test the nuclear TMD gluon density\cite{nLLM-2024} with available LHC data\cite{ChargedHadron5-CMS, ChargedHadron5-ATLAS, ChargedHadron5-ALICE}.
The consideration below continues the line of our studies\cite{LLM-2022, LLM-small-x-fit, LLM-2024, nLLM-2024}.

For the reader's convenience, first we recall some important formulas
for charged hadron production in $pp$ collisions.
It is known that the colliding protons at low scale $Q^2$ can be
considered as two systems consisting of three valence
quarks and a gluon environment with the wave function $|\Psi_g|^2 \sim f_g(x, {\mathbf k}_T^2)$,
where the gluon density $f_g(x, {\mathbf k}_T^2)$
depends on the proton longitudinal momentum fraction $x$, non-zero
transverse momentum ${\mathbf k}_T^2$
and does not depend on the scale $Q^2$ due to the gluon saturation effect at low $Q^2 \leq Q_s^2$,
where $Q^2_s$ is the saturation scale.
Then, the $pp$ interaction amplitude can be presented in the simple spectator form
$F_{pp} = f^{(0)}_{3q}\Psi_g$, where $f^{(0)}_{3q}$ is the amplitude of interaction of two $3q$ systems.
The cross section of a hadron $h$ production reads\cite{LLM-2022} (see also\cite{mQGSM-2,GLLZ})
\begin{gather}
    E {d^3 \sigma \over d^3 {\mathbf p}} \equiv {1\over \pi} {d^3 \sigma \over d^2 {\mathbf p}_T dy} = |F_{pp}|^2\otimes\left(D_{q/qq\rightarrow h} + D_{g\rightarrow h}\right) = \rho_q(x,p_T) + \rho_g(x,p_T),
    \label{def:rho_qg}
\end{gather}
\noindent
where the sign $\otimes$ means the convolution between $|F_{pp}|^2$ and fragmentation functions
$D_{q/qq\rightarrow h}$ and/or $D_{g\rightarrow h}$,
$E$, ${\mathbf p}$ and $y$ are the energy, $3$-dimensional momentum and
center-of-mass rapidity of the produced hadron having the transverse momentum $p_T \equiv |{\mathbf p}_T|$.
The quark/diquark contributions were calculated\cite{QGSM-1} within the
standard QGSM using only one-Pomeron exchange\footnote{In the midrapidity and
small $x_T = 2 p_T/\sqrt s$ the multi-Pomeron exchanges result in negligibly small contributions\cite{mQGSM-2,GLLZ}.}
and can be presented in the following form:
\begin{gather}
 \rho_q(x,p_T)=|f^{(0)}_{3q}|^2\otimes D_{q/qq\rightarrow h}\times\int d^2 {\mathbf k}_T dz f_g(z, {\mathbf k}_T^2) = \sigma_1 \phi_q(x,p_T)
    \label{def:rho_q}
\end{gather}
\noindent
with $\sigma_1$ being the cross section of one-Pomeron exchange (see\cite{QGSM-1} and references therein).
The gluon contribution can be calculated as similar convolution of the gluon distribution
with a fragmentation function $D_{g\rightarrow h}$
multiplied by $|f^{(0)}_{3q}|^2$ integrated over the intrinsic
phase space, which approximately results in the inelastic $pp$ cross-section $\sigma_{\rm in}$\cite{mQGSM-2,GLLZ}:
\begin{gather}
    \rho_g(x,p_T)=f_{g}\otimes D_{g\rightarrow h}\times\sigma_{\rm in} = \sigma_{\rm in} \phi_g(x,p_T).
    \label{def:rho_g}
\end{gather}
\noindent
The expressions for $\phi_q(x,p_T)$ and $\phi_g(x,p_T)$ can be written in the following form:
\begin{gather}
    \label{eq:phiq} \phi_q (x,p_T) = C_q \Big\{ \Phi_{q}(x_{+},p_T)\Phi_{qq}(x_{-},p_T) + \Phi_{qq}(x_{+},p_T)\Phi_{q}(x_{-},p_T) \Big\}, \\
    \label{eq:phig} \phi_g (x,p_T) = C_g \Big\{\Phi_g(x_+,p_T)+\Phi_g(x_-,p_T)\Big\},
\end{gather}
\noindent
where momentum fractions are
\begin{gather}
    x_{\pm} = {1\over 2} \left( \sqrt{x^2 + {4(m^2_h + p_T^2)\over s}} \pm x \right), \quad x = 2 \sqrt{m_h^2 + p_T^2 \over s} \sinh y,
    \label{eq:Kinematics1}
\end{gather}
\noindent
and $m_h$ being the produced hadron mass and $C_q$ and $C_g$ being the free parameters.
Here $\Phi_q(x, p_T)$, $\Phi_{qq}(x, p_T)$ and $\Phi_g(x, p_T)$ are the convolutions of
quark, diquark and gluon densities in a proton (derived with taking into account
energy-momentum conservation law) and corresponding
fragmentation functions into hadrons (namely, to pions and kaons).
Analytical expressions for $\Phi_q(x, p_T)$, $\Phi_{qq}(x, p_T)$ and $\Phi_g(x, p_T)$
are presented elsewhere\cite{LLM-2022}.
The inelastic $pp$ cross section involved in~(\ref{def:rho_g})
is calculated as difference between the total and elastic $pp$ scattering cross
sections: $\sigma_{\rm in} = \sigma_{\rm tot} - \sigma_{\rm el}$.
It can be parametrized as follows\cite{CrossSectionParametrization-1, CrossSectionParametrization-2}:
\begin{gather}
  \sigma_{\rm tot} = 21.7 (s/s_0)^{0.0808} + 56.08 (s/s_0)^{-0.4525}~{\rm mb},\\
  \sigma_{\rm el} = 11.84 - 1.617 \ln \left( s/s_0 \right) + 0.1359 \ln^2 \left(s/s_0 \right)~{\rm mb},
\end{gather}
\noindent
where $s_0 = 1$~GeV. The cross section of the one-Pomeron exchange $\sigma_1$
was calculated ealier\cite{QGSM-1}.
The fragmentation functions were calculated in the QCD at the leading (LO) and next-to-leading (NLO) orders\cite{QGSM-FFs}. So, for gluons one has
\begin{gather}
G_{g\to h}(z,|{\mathbf p}_T|) = 2 G_{g\to\pi}(z) I^g_{\pi}(|{\mathbf p}_T|) + 2 G_{g\to K}(z) I^g_{K}(|{\mathbf p}_T|),
\end{gather}
\noindent
where the coefficients $2$ come from the following relations:
\begin{gather}
G_{g\to\pi^+}(z) = G_{g\to\pi^-}(z), \quad G_{g\to K^+}(z) = G_{g\to K^-}(z).
\end{gather}
\noindent
The parametrizations of $G_{g\to \pi} (z)$ and $G_{g\to K} (z)$ are the following:
\begin{gather}
G_{g\to \pi} (z) = 6.57 z^{0.54}  (1-z)^{3.01}, \quad G_{g\to K} (z)  = 0.37 z^{0.79}  (1-z)^{3.07},
\end{gather}
\noindent
and functions $I^g_{\pi}(|{\mathbf p}_T|)$ and $I^g_{K}(|{\mathbf p}_T|)$ read:
\begin{equation}
I^g_{\pi}(|{\mathbf p}_T|) = I^g_{K}(|{\mathbf p}_T|) = I^g_{h}(|{\mathbf p}_T|) = \frac{(B^g_h)^2}{ 2\pi} e^{-B^g_h |{\mathbf p}_T|},
\label{Ig}
\end{equation}
\noindent
with $B^g_{h}$ being the free parameters which have to be determined from the data (see\cite{LLM-small-x-fit}).

Concerning the TMD gluon density, here we
follow our previous considerations\cite{LLM-2022, LLM-2024, nLLM-2024}.
So, in the case of proton, it was proposed in a simple form:
\begin{gather}
    f_g(x, {\mathbf k}_T^2) = c_g (1 - x)^{b_g} \sum_{n = 1}^3  c_n \left(R_0(x) |{\mathbf k}_T| \right)^n \exp( - R_0(x)|{\mathbf k}_T|), \nonumber \\
    R_0^2(x) = {1\over Q_0^2} \left({x\over x_0}\right)^{\lambda}, \quad b_g = b_g(0) + {4 C_A \over \beta_0} \ln {\alpha_s(Q_0^2) \over \alpha_s({\mathbf k}_T^2)},
    \label{eq-input}
\end{gather}
\noindent
where $C_A = N_c$, $\beta_0 = 11 - 2/3 N_f$, $Q_0 = 2.2$~GeV is the starting scale.
The corresponding saturation scale is given by $Q_s(x) = 1/R_0(x)$.
All phenomenological parameters, namely, $c_g$, $x_0$, $\lambda$, $b_g(0)$ and $c_n$
were determined\cite{LLM-small-x-fit, LLM-2024} from a fit to numerous HERA
and LHC data for some processes specially sensitive to the gluon content of a proton.
In particular, the gluon density (\ref{eq-input}) provides a self-consistent simultaneous description of the
HERA data on the proton structure functions $F_2(x,Q^2)$, $F_L(x,Q^2)$ and the reduced cross section $\sigma_r(x,Q^2)$
for electron-proton deep inelastic scattering at low $Q^2 \leq Q_0^2$.
Then, we have extended the expression (\ref{eq-input}) to nuclei\cite{nLLM-2024} using the empirical property of geometric scaling\cite{GeometryScaling-1}.
In fact, the HERA data on $\sigma^{\gamma^* p}(x,Q^2)$
can be described well by a single variable $\tau = Q^2/Q_s^2(x)$ in a wide range of $Q^2$,
where all $x$ dependence is encoded in the saturation scale $Q_s^2(x) \sim x^{- \lambda}$
with $\lambda \sim 0.3$. The same scaling effect was observed for
nuclear cross sections $\sigma^{\gamma^* A}(x,Q^2)$\cite{GeometryScaling-2}.
It was shown that
\begin{gather}
    {\sigma^{\gamma^* A}(\tau) \over \pi R_A^2} = {\sigma^{\gamma^* p}(\tau) \over \pi R_p^2},
    \label{eq-GeometryScalingA}
\end{gather}
\noindent
where $R_h$ is the radius of the hadronic target ($p$, $A$) and $\tau = \tau_A \equiv Q^2/Q_{s,\,A}^2$ or $\tau = \tau_p \equiv Q^2/Q_{s}^2$, respectively.
So, the $A$-dependence of the ratio $\sigma^{\gamma^* h}(x, Q^2) / \pi R_h^2$
can be absorbed in the $A$-dependence of $Q_{s,\,A}(x)$.
For this dependence, we have applied an ansatz\cite{GeometryScaling-2}:
\begin{gather}
    Q^2_{s,\,A}(x) = Q^2_s(x) \left( {A \pi R_p^2 \over \pi R_A^2} \right)^{1\over \delta}
    \label{eq-SaturationScaleA}
\end{gather}
\noindent
where $\delta$ and $\pi R_p^2$ are free parameters,
which were found\cite{nLLM-2024} by fitting of experimental data on ratios of nuclear structure functions $F_2^{A}(x,Q^2)/F_2^{A^\prime}(x,Q^2)$
at low $x$. Two well-known parametrizations of nucleus radius $R_A$, namely,
$R_A = \left(1.12 A^{1/3} - 0.86 A^{-1/3}\right)\text{\,fm}$\cite{GeometryScaling-2} (Fit I)
and $R_A = \left( 1.12 A^{1/3} - 0.5 \right)\text{\,fm}$\cite{RA-Textbook} (Fit II) have been used.
Then, the nuclear gluon distribution can be immediately obtained from the gluon density (\ref{eq-input}) after
replacement $Q^2_s(x) \to Q^2_{s,\,A}(x)$\cite{GeometryScaling-2} and applying the overall normalization factor $R_A^2/R_p^2$.

\begin{figure}
	\begin{center}
	  \includegraphics[width=5.20cm]{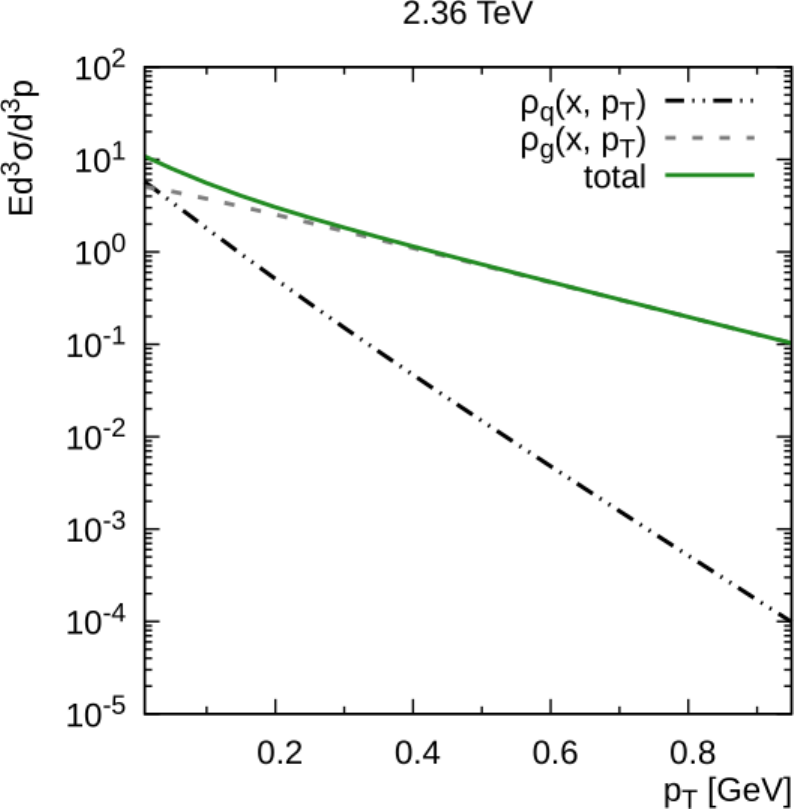}
	  \includegraphics[width=5.20cm]{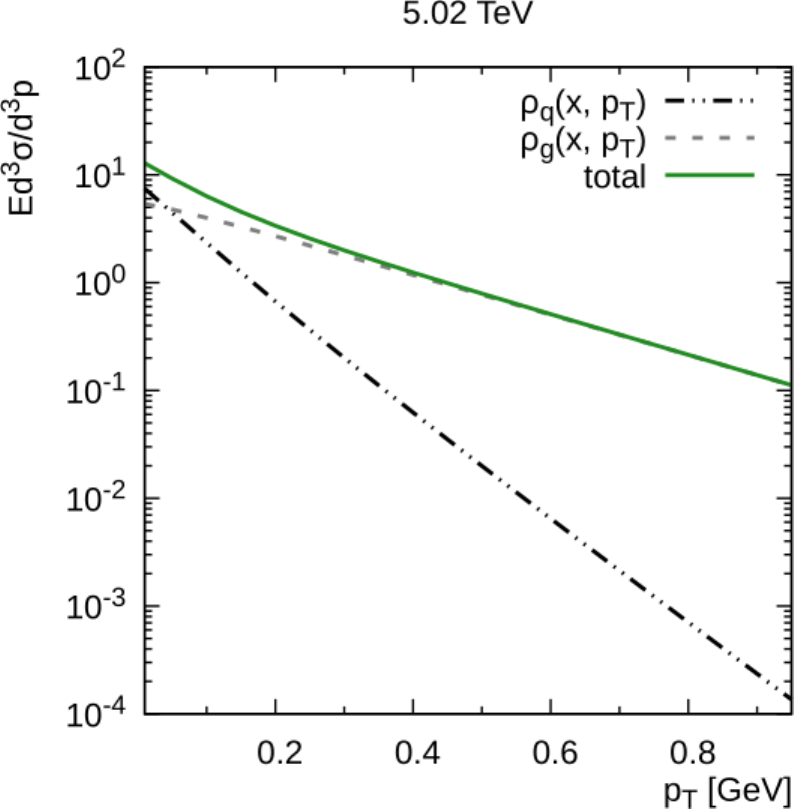}
	  \includegraphics[width=5.20cm]{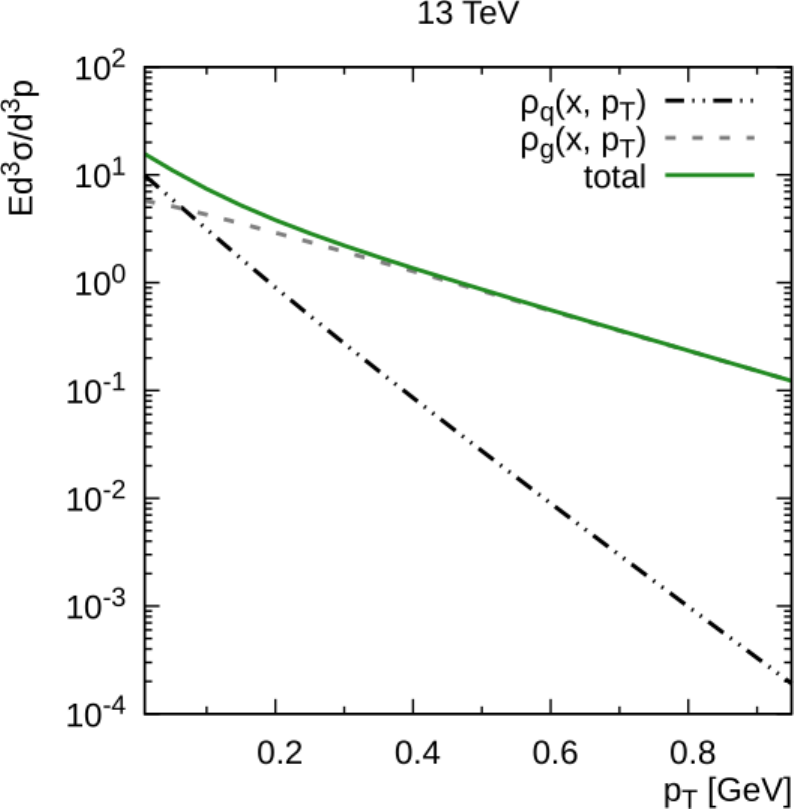}
	  \caption{Quark/diquark (black dash-dotted) and gluon (gray dashed)
          contributions to the inclusive spectrum of soft charged hadrons 
          calculated at $\sqrt{s} = 2.36$, $5.02$ and $13$~TeV. Sum of all contributions is represented by the green line.}
		\label{fig1}
	\end{center}
\end{figure}

Now we can easily extend the formalism of the modified QGSM to $pA$ interactions.
To do it, we note that both quark/diquark and gluon contributions
(\ref{def:rho_q}) and (\ref{def:rho_g}) to hadron production cross section
are related to the gluon density $f_g(x, {\mathbf k}_T^2)$.
However, terms depending on momentum fraction $x_-$ in (\ref{eq:phiq}) and (\ref{eq:phig})
correpond to the fragmentation of quarks/diquarks/gluons in the proton into hadrons
whereas the terms depending on $x_+$ variable represent the quarks/diquarks/gluons fragmentation
in the nucleus $A$.
Therefore, the geomeric scaling described above should be
applied to these terms only with keeping the
general expression ({\ref{def:rho_qg}}) for hadron spectra unchanged.
Note that the quark/diquark contribution to the the inclusive hadron spectra
at $p_T \geq 0.2$~GeV is very small and can be safely neglected as
the gluons dominate at the LHC energies.
It is clearly illustrated in Fig.~\ref{fig1}
for $\sqrt s = 2.36$, $5.02$ and $13$~TeV.
Thus, in the calculations below we omit
quark/diquark terms and do not extend them to the $pA$ collisions.

As we transfer from the proton to the nucleus, one has to consider that the parameters of the gluon fragmentation function can alter. It is known that the jet quenching effect leads to a softening of particle production spectra in nuclear collisions\cite{jetquenching1,jetquenching2}. Therefore we perform fits on $\pi$ and $K$-meson production data to take into account such effects. Concerning the $K$-meson production data, the slope parameter $B^g_K$ appearing
in~(\ref{eq-input}) was taken\cite{LLM-2022,LLM-small-x-fit} to be equal to the one of $\pi$-mesons. The reason for this was that in the considered $pp$-data the spectra were presented as sums of the mesons production. Since the $\pi$-mesons yield was dominant, it was hard to extract the $K$-meson contribution from the data. Here we will determine both $B^g_\pi$ and $B^g_K$ independently.



\begin{figure}
    \begin{center}
        \includegraphics[width=5.15cm]{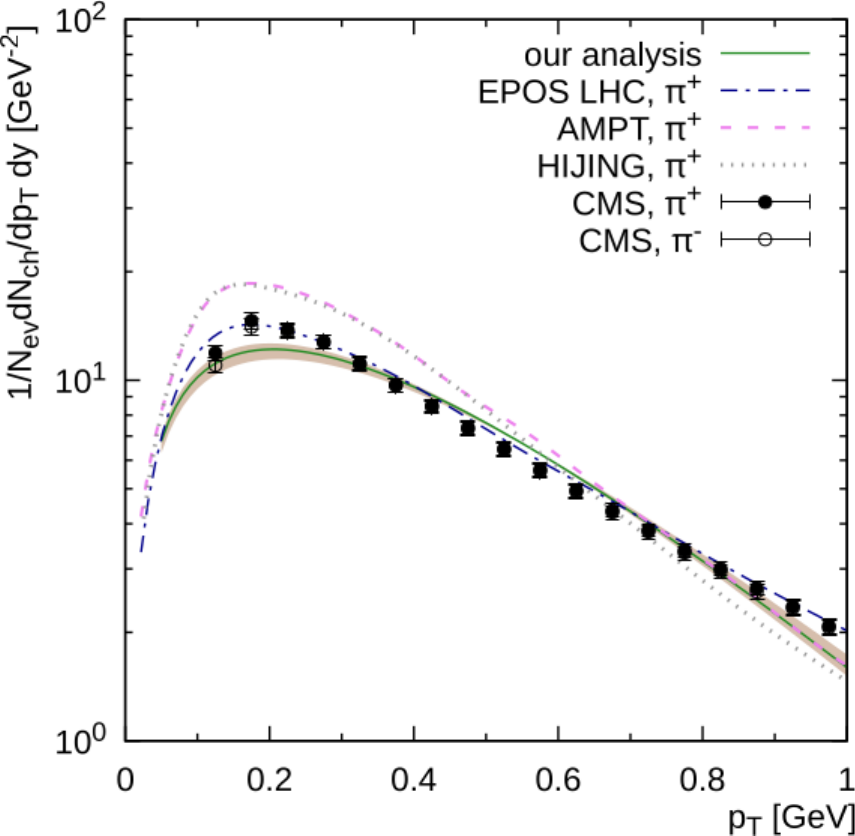}
        \includegraphics[width=5.3cm]{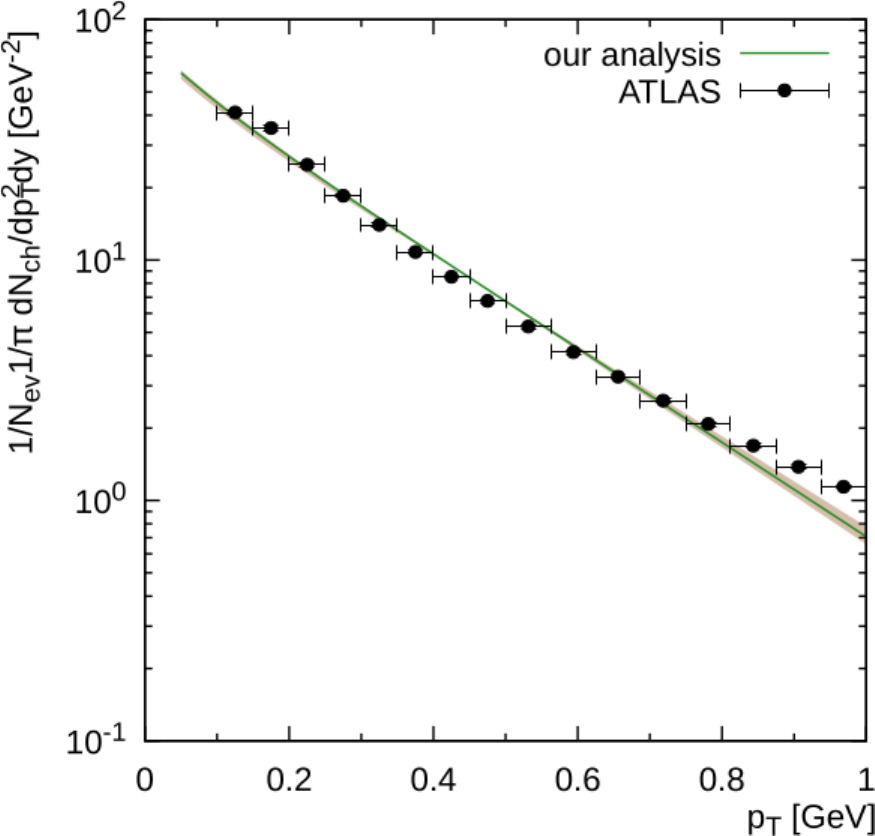}
        \includegraphics[width=5.2cm]{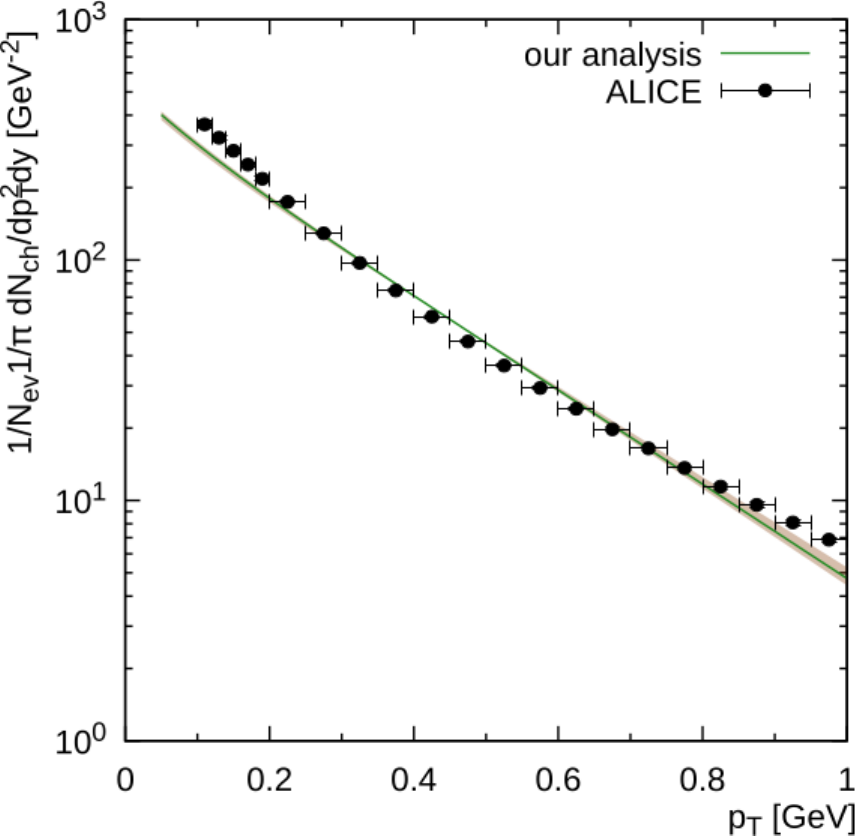}
        \caption{Pions transverse momenta spectra calculated within our approach compared with available LHC data
                for $p_T < 1$~GeV.
                The uncertainties of the fit procedure are shown as brown bands.
                The data of CMS~\cite{ChargedHadron5-CMS}, ATLAS~\cite{ChargedHadron5-ATLAS} and ALICE~\cite{ChargedHadron5-ALICE} were taken at $\sqrt{s}=$ 5.02~TeV and summed over the collision centralities.}
        \label{fig2}
    \end{center}
\end{figure}

\begin{figure}
    \begin{center}
        \includegraphics[width=5.2cm]{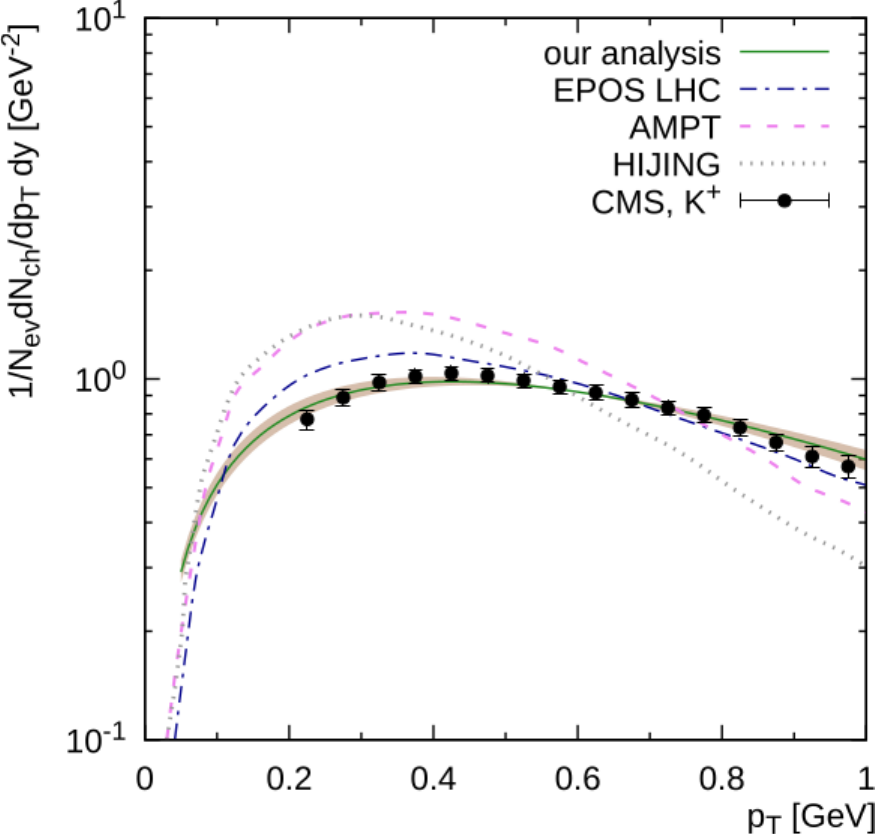}
        \includegraphics[width=5.2cm]{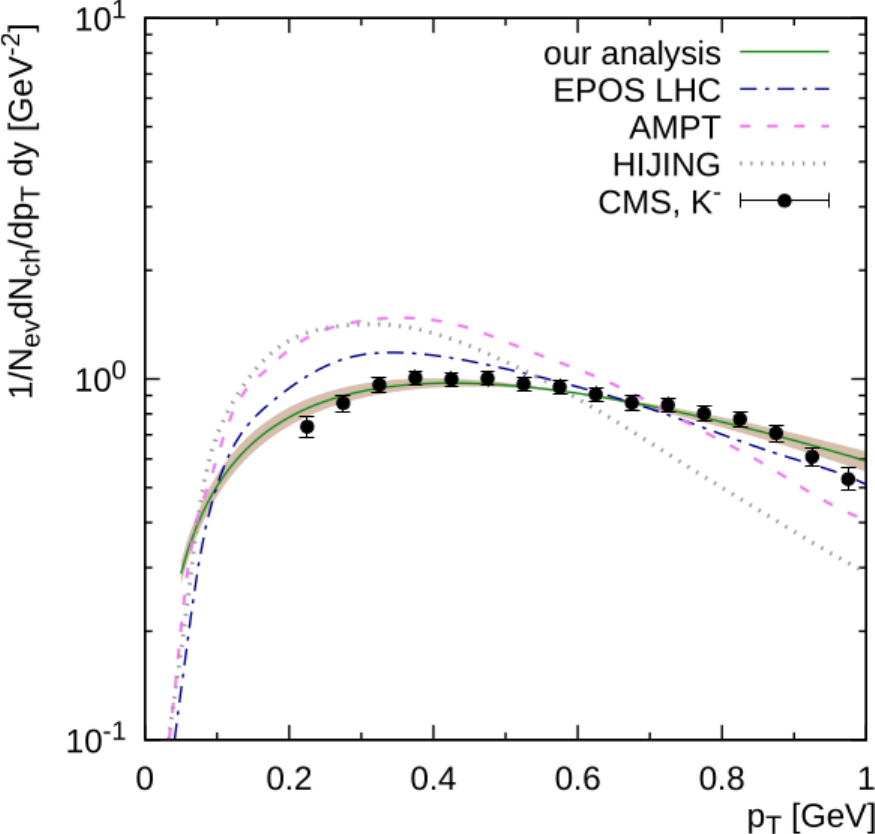}
        \includegraphics[width=5.1cm]{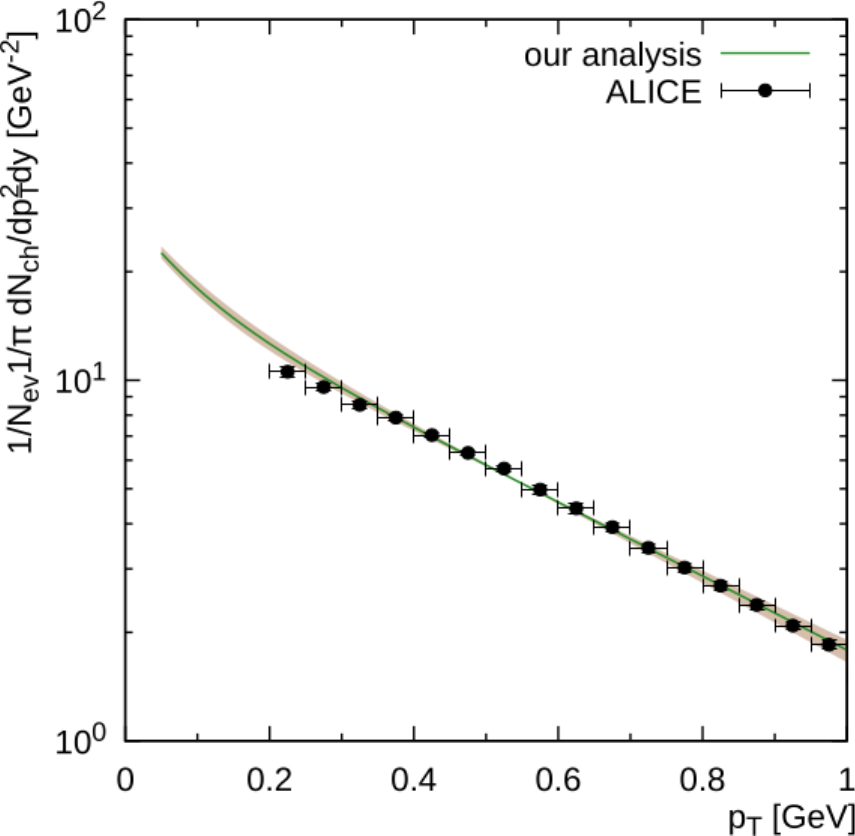}
        \caption{Kaons transverse momenta spectra calculated within our approach compared with available LHC data
                for $p_T < 1$~GeV.
                The uncertainties of the fit procedure are shown as brown bands.
                The data of CMS~\cite{ChargedHadron5-CMS}, ATLAS~\cite{ChargedHadron5-ATLAS} and ALICE~\cite{ChargedHadron5-ALICE} were taken at $\sqrt{s}=$ 5.02~TeV and summed over the collision centralities.}
        \label{fig3}
    \end{center}
\end{figure}

  Our results for the transverse momentum distributions of inclusive charged hadrons (pions and kaons)
  produced in proton-lead collisions at $\sqrt{s} = 5.02$~TeV compared to CMS\cite{ChargedHadron5-CMS}, ATLAS\cite{ChargedHadron5-ATLAS} and ALICE\cite{ChargedHadron5-ALICE} data are presented in Figs.~\ref{fig2} and \ref{fig3}. Note that we are constrained by our model to the central region, so among the ATLAS data we choose ones taken at the center-of-mass rapidity $|y^*|<0.5$. The CMS data were taken at the laboratory rapidity $|y|<1$ and ALICE made measurements at $0<|y^*|<1$. The data we use are summed over the centralities of the collisions
  and refers to small $p_T$ region, $p_T < 1$~GeV.
 Our predictions are depicted with solid curves, whereas brown bands represent
 the uncertainty of the $B^g_\pi$ and/or $B^g_\pi$ fits, respectively.
  Note that the different formulae for the nuclear radius $R_A$ mentioned above coincide at large $A=208$, so that numerically
 we took first of them (Fit~I) for definiteness. We find that best description of proton-lead data on $\pi$-mesons production is
 achieved with $B^g_\pi=9.1^{+0.2}_{-0.5}$. For $pp$ collisions, we had $B^g_\pi=8.7\pm 0.3$\footnote{In\cite{LLM-small-x-fit} values twice as low were reported mistakenly.}. The obtained values coincide within uncertainties in fact. As a result, one can see a good agreement with the LHC data. A slight deviation from the data points can be only seen for $p_T\sim 1$ for ALICE and ATLAS results. This may indicate the necessity to include
 the QCD evolution effects already at such scales.
 Note that ALICE data also reveal some disagreement at very low $p_T < 0.2$~GeV,
 that might point to a lack of quark/diquark contributions.
 For $K$-mesons, we have obtained $B^g_K=4.8^{+0.4}_{-0.3}$, that results in an excellent agreement with the available data.
Additionally, we show the predictions from three Monte-Carlo event generators,
namely, \textsc{hijing}\cite{HIJING}, \textsc{ampt}\cite{AMPT} and \textsc{epos lhc}\cite{EPOS} (we took all of them from the CMS paper\cite{ChargedHadron5-CMS}). It was claimed
that \textsc{epos lhc} reproduces well the measured hadron spectra, whereas \textsc{ampt} and \textsc{hijing}
predict steeper $p_T$ distributions and much smaller average $p_T$
than obtained in the CMS experiment\cite{ChargedHadron5-CMS}.
One can see that our results for $\pi^\pm$ production are very close to the \textsc{epos lhc} ones
within the fit uncertainties and agree well with the LHC data.
The achieved description is better than agreement provided by the \textsc{ampt} and \textsc{hijing} tools.
Moreover, our calculations for kaons leads to a best description of the CMS data.
It demonstrates the possibility of developed approach and
indicates that nuclear extension\cite{nLLM-2024} of proposed TMD gluon density\cite{LLM-2024} does not contradict
the available collider data on $pA$ events.

To summarize, in this paper we have extended the
modified quark-gluon string model to $pA$ interactions and
investigated the nuclear TMD gluon density proposed earlier\cite{nLLM-2024} for low $x$.
The application of developed formalism to the description of transverse momentum distributions
of soft pions and kaons produced inclusively in proton-lead collisions
results in a satisfactory coincidence with available CMS, ATLAS and ALICE data
collected at $\sqrt s = 5.02$~TeV.
Moreover, it provides, in general, a somewhat better description of low-$p_T$ data
than available tools (such as Monte-Carlo event generators \textsc{epos lhc}, \textsc{hijing} and \textsc{ampt}).
This result can be considered as an additional validation of nTMD gluon distribution
at low $x$ and not large hard scale ($\sim 1$~GeV), additionally to nuclear shadowing
effects analyzed earlier\cite{nLLM-2024}.

{\sl Acknowledgements.} We thank S.P.~Baranov and H.~Jung for their
interest, very important comments and remarks.
This research has been carried out at the expense of the Russian Science Foundation grant No.~25-22-00066, https://rscf.ru/en/project/25-22-00066/.


\begin{thebibliography}{10}

\bibitem{NuclReview-1}
M.~Arneodo{,} Phys. Rept. {\bf 240}{,}~301 (1994).

\bibitem{NuclReview-2}
P.R.~Norton{,} Rept. Prog. Phys. {\bf 66}{,}~1253 (2003).

\bibitem{NuclReview-3}
S.~Malace{,} D.~Gaskell{,} D.W.~Higinbotham{,} I.C.~Cl\"ot{,} Int. J. Mod.
  Phys. E {\bf 23}{,}~1430013 (2014).

\bibitem{NuclearMomentumDistribution}
W.B.~Atwood{,} G.B.~West{,} Phys. Rev. D {\bf 7}{,}~773 (1973).

\bibitem{NuclearBindingCorrections}
S.V.~Akulinichev{,} S.A.~Kulagin{,} G.M.~Vagradov{,} Phys. Lett. B {\bf
  158}{,}~485 (1985).

\bibitem{NuclearShadowingApproach-1}
B.Z.~Kopeliovich{,} A.~Sch\"afer{,} A.V.~Tarasov{,} Phys. Rev. D {\bf
  62}{,}~054022 (2000).

\bibitem{NuclearShadowingApproach-2}
J.-W.~Qiu{,} I.~Vitev{,} Phys. Lett. B {\bf 632}{,}~507 (2006).

\bibitem{ChargedHadron5-CMS}
CMS Collaboration{,} Eur. Phys. J. C {\bf 74}{,}~2847 (2014).

\bibitem{ChargedHadron5-ATLAS}
ATLAS Collaboration{,} Phys. Lett. B {\bf 763}{,}~313 (2016).

\bibitem{ChargedHadron5-ALICE}
ALICE Collaboration{,} Phys. Lett. B {\bf 728}{,}~25 (2014).

\bibitem{EPOS}
T.~Pierog{,} Iu.~Karpenko{,} J.M.~Katzy{,} E.~Yatsenko{,} K.~Werner{,} Phys.
  Rev. C {\bf 92}{,}~034906 (2015).

\bibitem{AMPT}
Z.W.~Lin{,} Indian J. Phys. {\bf 85}{,}~837 (2011).

\bibitem{HIJING}
W.-T.~Deng{,} X.-N.~Wang{,} R.~Xu{,} Phys. Rev. C {\bf 83}{,}~014915 (2011).

\bibitem{mQGSM-2}
V.A.~Bednyakov{,} A.A.~Grinyuk{,} G.I.~Lykasov{,} M.~Poghosyan{,} Int. J. Mod.
  Phys. A {\bf 27}{,}~1250042 (2012).

\bibitem{GLLZ}
A.A.~Grinyuk{,} A.V.~Lipatov{,} G.I.~Lykasov{,} N.P.~Zotov{,} Phys. Rev. D {\bf
  87}{,}~074017 (2013).

\bibitem{QGSM-1}
A.B.~Kaidalov{,} Z. Phys. C {\bf 12}{,} 63 (1982);\\ A.B.~Kaidalov{,} Surveys
  High Energy Phys. {\bf 13}{,} 265 (1999);\\ A.B.~Kaidalov{,}
  O.I.~Piskunova{,} Z. Phys. C {\bf 30}{,}~145 (1986).

\bibitem{LLM-2022}
A.V.~Lipatov{,} G.I.~Lykasov{,} M.A.~Malyshev{,} Phys. Rev. D {\bf
  107}{,}~014022 (2023).

\bibitem{LLM-2024}
A.V.~Lipatov{,} G.I.~Lykasov{,} M.A.~Malyshev{,} JETP Lett. {\bf 119}{,}~828
  (2024).

\bibitem{HighEnergyFactorization}
S.~Catani{,} M. Ciafaloni{,} F. Hautmann{,} Nucl. Phys. B {\bf 366}{,} 135
  (1991){;} \\ J.C. Collins{,} R.K. Ellis{,} Nucl. Phys. B {\bf 360}{,}~3
  (1991).

\bibitem{kt-factorization}
L.V. Gribov{,} E.M. Levin{,} M.G. Ryskin{,} Phys. Rep. {\bf 100}{,} 1 (1983){;}
  \\ E.M. Levin{,} M.G. Ryskin{,} Yu.M. Shabelsky{,} A.G. Shuvaev{,} Sov. J.
  Nucl. Phys. {\bf 53}{,}~657 (1991).

\bibitem{TMD-review}
R.~Angeles-Martinez{,} A. Bacchetta{,} I.I. Balitsky{,} D. Boer{,} M.
  Boglione{,} R. Boussarie{,} F.A. Ceccopieri{,} I.O. Cherednikov{,} P.
  Connor{,} M.G. Echevarria{,} G. Ferrera{,} J. Grados Luyando{,} F.
  Hautmann{,} H. Jung{,} T. Kasemets{,} K. Kutak{,} J.P. Lansberg{,} A.
  Lelek{,} G.I. Lykasov{,} J.D. Madrigal Martinez{,} P.J. Mulders{,} E.R.
  Nocera{,} E. Petreska{,} C. Pisano{,} R. Placakyte{,} V. Radescu{,} M.
  Radici{,} G. Schnell{,} I. Scimemi{,} A. Signori{,} L. Szymanowski{,} S.
  Taheri Monfared{,} F.F.~Van der Veken{,} H.J.~van Haevermaet{,} P. Van
  Mechelen{,} A.A. Vladimirov{,} S. Wallon{,} Acta Phys. Polon. B {\bf 46}{,}
  2501~(2015).

\bibitem{TMD-review-our}
A.V.~Lipatov{,} S.P.~Baranov{,} M.A.~Malyshev{,} Phys. Part. Nucl. {\bf
  55}{,}~256 (2024).

\bibitem{LLM-small-x-fit}
A.V.~Lipatov{,} G.I.~Lykasov{,} M.A.~Malyshev{,} Phys. Lett. B {\bf
  848}{,}~138390 (2024).

\bibitem{nLLM-2024}
A.V.~Lipatov{,} G.I.~Lykasov{,} M.A.~Malyshev{,} Phys. Rev. D {\bf
  112}{,}~014025 (2025).

\bibitem{GeometryScaling-1}
A.M.~Stasto{,} K.~Golec-Biernat{,} J.~Kwiecinski{,} Phys. Rev. Lett. {\bf
  86}{,}~596 (2001).

\bibitem{GeometryScaling-2}
N.~Armesto{,} C.A.~Salgado{,} U.A.~Wiedemann{,} Phys. Rev. Lett. {\bf
  94}{,}~022002 (2005).

\bibitem{CrossSectionParametrization-1}
N.~Cartiglia{,} arXiv:1305.6131 [hep{-}ph].

\bibitem{CrossSectionParametrization-2}
I.M.~Dremin{,} Particles {\bf 2}{,}~57 (2019).

\bibitem{QGSM-FFs}
J.~Binnewies{,} B.A.~Kniehl{,} G.~Kramer{,} Phys. Rev. D {\bf 52}{,}~4947
  (1995).

\bibitem{RA-Textbook}
I.M.~Kapitonov{,} "An~Introduction to~Physics~of Nuclei and Particles"{,}
  URSS{,}~Moscow{,} 2002.

\bibitem{jetquenching1}
S.~Cao{,} X.-N. Wang{,} Rept. Prog. Phys. {\bf 84}{,}~024301 (2021).

\bibitem{jetquenching2}
N.~Armesto{,} L. Cunqueiro{,} C.A. Salgado{,} W.-C. Xiang{,} JHEP {\bf
  02}{,}~048 (2008).

\end{thebibliography}


\end{document}